\documentclass[useAMS,usenatbib]{mn2e}
\usepackage{aas_macros,graphicx,times,multirow,amsmath}
\usepackage[]{color}

\title[Swift monitoring of the X-ray source in \snr]{\swift\ monitoring of the
central X-ray source in RCW\,103}
\author[P. Esposito et al.]{P.~Esposito,$^{1}$\thanks{E-mail: paoloesp@oa-cagliari.inaf.it}
R.~Turolla,$^{2,3}$
A.~De~Luca,$^{4,5}$ G.~L.~Israel,$^{6}$ A.~Possenti$^{1}$ and D.~N.~Burrows$^{7}$
\smallskip\\
$^1$INAF -- Osservatorio Astronomico di Cagliari, localit\`a Poggio dei Pini, strada 54, I-09012 Capoterra, Italy\\
$^2$Universit\`a di Padova, Dipartimento di Fisica, via F.~Marzolo 8, I-35131 Padova, Italy\\
$^3$Mullard Space Science Laboratory, University College London, Holmbury St. Mary, Dorking, Surrey RH5 6NT, UK\\
$^4$IUSS -- Istituto Universitario di Studi Superiori, viale Lungo Ticino Sforza 56, I-27100 Pavia, Italy\\
$^5$INAF -- Istituto di Astrofisica Spaziale e Fisica Cosmica - Milano, via E. Bassini 15, I-20133 Milano, Italy\\
$^6$INAF -- Osservatorio Astronomico di Roma, via Frascati 33, I-00040 Monteporzio Catone, Italy\\
$^7$Department of Astronomy \& Astrophysics, The Pennsylvania State University, 525 Davey Laboratory,
University Park, PA 16802, USA}
\date{Accepted 2011 July 19.  Received 2011 July 19; in original form 2011 July 5} 
\pagerange{\pageref{firstpage}--\pageref{lastpage}} \pubyear{2011}

\def\LaTeX{L\kern-.36em\raise.3ex\hbox{a}\kern-.15em
    T\kern-.1667em\lower.7ex\hbox{E}\kern-.125emX}

\def\xmm {\emph{XMM-Newton}}
\def\cxo {\emph{Chandra}}
\def\swift {\emph{Swift}}

\def\src {1E\,1613}
\def\snr {RCW\,103}
\def\flux {\mbox{erg cm$^{-2}$ s$^{-1}$}}

\begin{document}

\label{firstpage}
\maketitle
\begin{abstract}

The X-ray source 1E\,161348--5055 lies at the centre of the 2-kyr-old supernova remnant \snr. Owing to its 24-ks modulation, orders-of-magnitude flux variability over a few months/years, and lack of an obvious optical counterpart, 1E\,161348--5055 defies assignment to any known class of X-ray sources. Starting from April 2006, \swift\ observed 1E\,161348--5055 with its X-ray telescope for $\sim$2 ks approximately once per month. During the five years covered, the source has remained in a quiescent state, with an average observed flux of $\sim$$1.7\times10^{-12}$ \flux\ (1--10 keV), $\sim$20 times lower than the historical maximum attained in its 1999--2000 outburst. The long time-span of the \swift\ data allows us to obtain an accurate measure of the period of 1E\,161348--5055 [$P=24\,030.42(2)$ s] and to derive the first upper limit on its period derivative ($|\dot{P}|<1.6\times10^{-9}$ s s$^{-1}$ at 3$\sigma$).
\end{abstract}

\section{Introduction}

1E\,161348--5055 (hereafter \src) was discovered with the \emph{Einstein} satellite \citep{tuohy80} close to the geometrical centre of the young supernova remnant (SNR) \snr\ (age $\sim$2 kyr; \citealt*{carter97}). It was proposed as the first example of a radio-quiet (possibly owing to an unfavourable radio beaming), isolated, cooling neutron star (\citealt{tuohy80,tuohy83}; \citealt*{gph97}).

At present, there is little doubt that \src\ indeed is a neutron star \citep{deluca06} and
the source is traditionally included in the class of the `central compact objects' (CCOs; see
\citealt{deluca08} for a review). CCOs are a small group of young and seemingly isolated
X-ray-emitting neutron stars (with thermal-like spectra), observed close to the centre of
non-plerionic SNRs and without obvious counterparts in other wavebands. However, its peculiar
temporal behaviour distinguishes \src\ from the other CCOs (actually, it singles this source
out as a unique object in general). The first peculiarity of the source is its
orders-of-magnitude X-ray flux variability on a few months/years time-scale
(\citealt*{gotthelf99}; \citealt{garmire00,sanwal02,becker02}).
Moreover, the first \cxo\ observation of \src\
in a low state hinted at a possible periodicity at $\sim$6 hours \citep{gpg00} that was not
confirmed by subsequent observations of the source in bright states. A long (90 ks)
observation with \xmm, performed in 2005, caught \src\ in a low state and yielded unambiguous
evidence for a strong, nearly sinusoidal modulation at $6.67\pm0.03$ hours ($24.0\pm0.1$
ks; \citealt{deluca06}). The same periodicity was then
recognised also in the older data-sets, albeit with a very different pulse shape,
including two narrow dips per period.
No faster pulsations are seen in \src\ \citep{deluca06}.

Large flux variations, similar to those observed in \src, are common among magnetars (e.g.
\citealt{rea11}), but these pulsars, whose emission is believed to be powered mainly by the
magnetic field, are characterised by rotational periods in the narrow range 2--12 s. On the
other hand, CCOs are steady sources, and their periods --when known-- are in the 0.1--0.5 s
range (\citealt{zavlin00}; \citealt*{ghs05,gotthelf09}). If \src\ is indeed a magnetar, it must have been
slowed down by some unusual mechanisms, perhaps by a propeller interaction with a debris disk
\citep{deluca06,li07}. A different possibility is that \src\ is a peculiar low-mass
binary,\footnote{Deep observations of the field of \src\ with the Very Large Telescope and
the \emph{Hubble Space Telescope} showed only two or three faint infrared sources ($H\sim22$)
consistent with the position of \src. If none of them is linked to the X-ray
source, \src\ is undetected in the near infrared down to $H>23$ \citep{dmz08}.} powered by a double (wind
plus disk) accretion onto a recently-born compact object (in this case the 24-ks signal would
result from the orbital motion of the system) or hosting a magnetar
\citep{deluca06,pizzolato08,bhadkamkar09}. Both scenarios require nonstandard assumptions
about the formation and evolution of compact objects in supernova explosions.

Here we report on the results from a 5-year monitoring of \src\ with the \swift\ satellite
\citep{gehrels04short}. This unique data set allowed us to obtain a phase-coherent timing
solution encompassing also \cxo\ and \xmm\ archival observations. Thanks to this, we are able
to derive an accurate period for \src\ and to set the first upper limits on the period
derivative for this puzzling source.

\section{X-ray observations and data reduction}

\subsection{\swift\ data}

The X-Ray Telescope (XRT; \citealt{burrows05short}) on-board \swift\ uses a front-illuminated
CCD detector sensitive to photons between 0.2 and 10 keV with an effective area of about 110
cm$^2$ (at 1.5 keV) and a field view of 23-arcmin in diameter. Two main readout modes are
available: photon counting (PC) and windowed timing (WT). PC mode provides two dimensional
imaging information and a 2.5073-s time resolution; in WT mode only one-dimensional imaging is
preserved, achieving a time resolution of 1.766 ms.

\begin{figure} \resizebox{\hsize}{!}{\includegraphics[angle=-90]{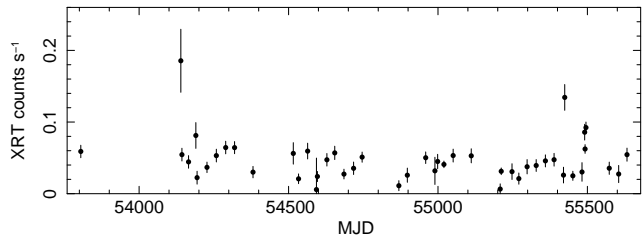}}
\caption{\label{swiftlc} \swift/XRT light curve of \src\ (1--10 keV). Each
observation is represented by a point in the plot. A rough conversion of 0.1
count s$^{-1}\simeq5\times10^{-12}$ \flux\ (1--10 keV, not corrected for absorption) can
be derived from the spectral analysis (Section~\ref{analysis}).}
\end{figure}

\begin{figure} \resizebox{\hsize}{!}{\includegraphics[angle=0]{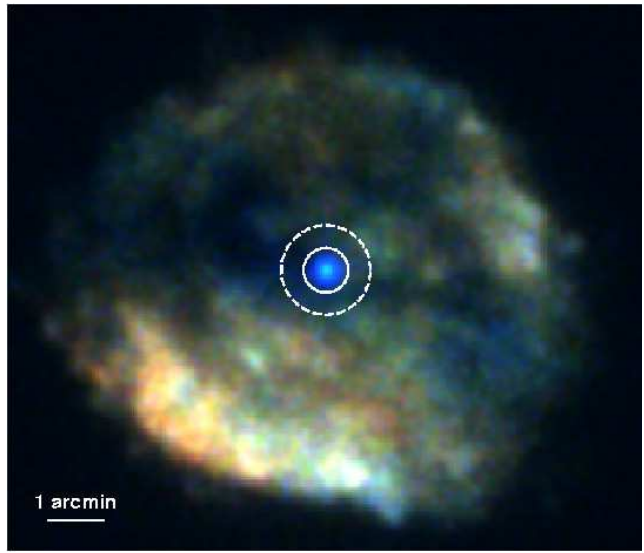}}
\caption{\label{ds9} \swift/XRT image of \src\ and its SNR, RCW\,103. All the
XRT data (as of 2011 April) were combined, totalling to 102.8 ks exposure time. Photon energy
is colour-coded: red corresponds to 0.2--0.9 keV energies, green to 0.9--1.7 keV, blue
to 1.7--8.0 keV. North is up, east is left. The source and background regions considered for
the analysis (Section~\ref{analysis}) are over-plotted.}
\end{figure}

Between April 2006 and April 2011, \src\ was observed by XRT 49 times, for a total net
exposure time of 102.8 ks in PC mode.\footnote{ Also 5.5 ks of WT data were collected during
the same pointings. However, given the bright SNR in which \src\ is embedded
(see Fig.~\ref{ds9}), we did not make use of them in this work.}
The distribution of the \swift\ observations can be seen in the long-term light curve in
Fig.~\ref{swiftlc}, while in Fig.~\ref{ds9} we show the image of \src\ and \snr\ resulting
from all the XRT data gathered so far. Except for periods in which the
source was not visible by the XRT because of pointing constraints of the \swift\ spacecraft,
approximately one 2-ks observation in imaging mode was collected per month. On a few occasions,
when \src\ showed hints of consistent flux variations, we requested target-of-opportunity
observations. For example this happened at the end of October 2010 (see Fig.~\ref{swiftlc}
around MJD 55\,500), when the source count rate remained at a relatively high level of
$\approx$0.09 counts s$^{-1}$ for a few consecutive pointings spanning $\sim$5 days.
The observations were not time-constrained, so the monitoring can be considered a
\emph{casual} sampling of the phase of \src.

The XRT data were uniformly processed with \textsc{xrtpipeline} (version 12, in the
\textsc{heasoft} software package version 6.9), filtered and screened with standard
criteria. In order to reduce the contamination from the SNR, the source counts were
energy-selected in the 1--10 keV band and extracted within a 10-pixel radius (one XRT pixel
corresponds to about $2\farcs36$). To convert the photon arrival times to the Solar system
barycentre for the timing analysis, we used the \textsc{barycorr} task and the \cxo\
position of \citet{dmz08}. For the spectroscopy, we used the latest spectral redistribution
matrices in \textsc{caldb} (20091130), while the ancillary response files were generated with
\textsc{xrtmkarf}, which accounts for different extraction regions, vignetting and
point-spread function corrections.

\subsection{\cxo\ and \xmm\ data} We complemented the \swift/XRT data-set with the few \cxo\
and \xmm\ observations long-enough to contain a minimum of two modulation cycles of \src\
(Table~\ref{obs-log}). The \cxo/ACIS-S observation (performed on 2002 March 03, when
the source was rather bright) has been already published in \citet{sanwal02}, the \xmm/EPIC
(2005 August 23--24) and \cxo/HRC-S (2007 July 03) ones, both carried out while \src\ was in a
low state, have been published in \citet{deluca06} and in \citet{dmz08}, respectively; we
refer to these papers for more details. The \cxo/ACIS-I observation of 2010 June 01 is reported
here for the first time. For this work the data were processed and analysed
with standard procedures, using the latest available versions of the \cxo\ Interactive
Analysis of Observation software (\textsc{ciao}, version 4.2) and of the \xmm\ Science
Analysis Software (\textsc{sas}, version 10).

\begin{table}
\centering
\caption{\xmm\ and \cxo\ observations used for this work.}
\label{obs-log}
\begin{tabular}{@{}cccc}
\hline
Instrument & Obs.ID & Date$^{a}$ & Duration$^{b}$\\
 & & (MJD TBD) & (ks)\\
\hline
\cxo/ACIS-S & 2759 & 52336.489 & 50.3 \\
\xmm/EPIC & 0302390101 & 53605.824 & 87.5 \\
\cxo/HRC-S & 7619 & 54284.330 & 80.2 \\
\cxo/ACIS-I & 11823 & 55348.604 & 62.5 \\
\hline
\end{tabular}
\begin{list}{}{}
\item[$^{a}$] Mid-point of observation.
\item[$^{b}$] Time between first and last event.
\end{list}
\end{table}

\section{Analysis and results}\label{analysis}

We merged the data from the \swift/XRT observations and accumulated a combined spectrum (a
detailed spectral analysis will be reported elsewhere). The data were rebinned with a minimum
of 20 counts per energy bin and the background was estimated from an annular region centred on
\src\ (with radii 10 and 20 pixel, see Fig.~\ref{ds9}). The spectrum can be fit
[$\chi^2_\nu=0.82$ for 174 degrees of freedom (dof)] by a double-blackbody corrected for the
interstellar absorption (see \citealt{deluca06}).
The best-fit parameters are blackbody temperatures $kT_1=0.50^{+0.06}_{-0.11}$ keV and
$kT_2=0.8^{+0.5}_{-0.1}$ keV, radii $R_1=0.6^{+0.2}_{-0.1}$ km and $R_2=0.12^{+0.10}_{-0.09}$ km
(for a distance of 3.3 kpc; \citealt{caswell75}), and absorption $N_{\rm H}=(7^{+1}_{-2})\times10^{21}$
cm$^{-2}$ (1$\sigma$ errors). The observed averaged flux is
$\sim$$1.7\times10^{-12}$ \flux, similar to that measured with \xmm\ in the August 2005
\citep{deluca06}.

As can be seen from Fig.~\ref{swiftlc}, \src\ showed only moderate
variability between the many \swift\ pointings, remaining always well below the flux level of
the 1999--2000 outburst ($\sim$$5\times10^{-11}$ \flux; \citealt{garmire00}). Also the \cxo\
HRC-S and ACIS-I observations show a flux of $\approx$$2\times10^{-12}$ \flux.
We note that on a statistical ground we are sensitive only to outbursts
lasting $\ga$ 1 month. Although shorter phases of enhanced emission cannot be
ruled out for \src\ based on our data, the historical behaviour of the
source indicates that its outbursts are likely longer than a few months
\citep{deluca06}. Thus our data suggest that  \src\ persisted in a quiescent state
during the five years of the \swift\ monitoring.

Owing to their long time span (from MJD 53\,804.500 to 55\,632.581), the \swift\ data are
suitable for studying the 24-ks modulation of \src. A folded profile showing a significant
modulation cannot be obtained from the XRT data using the most accurate period
available so far, i.e. the one estimated from
the August 2005 \xmm\ observation ($P=24.0\pm0.1$ ks; \citealt{deluca06}). This is not
surprising since $\sim$200 days separate the \xmm\ observation from the start of the \swift\
monitoring, and the \xmm\ period uncertainty implies a phase uncertainty of half a cycle after
only $\sim$30 days.

\begin{figure} \resizebox{\hsize}{!}{\includegraphics[angle=0]{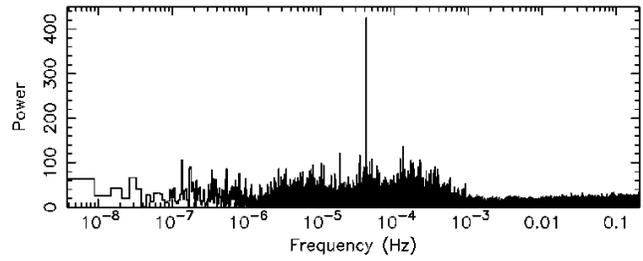}}
\caption{\label{powspec} Power spectrum computed for the combined \swift/XRT data. The
peak corresponding to the 24-ks signal is evident.} \end{figure}

So, as a starting point, we computed a fast-Fourier-transform power spectrum using all
the \swift\ data at the highest resolution allowed by the PC mode (bin time 2.5073 s).
Given the approximate knowledge of the source period, a `blind' search is not, in principle,
necessary, but we did this to have a clear picture of the \swift\ time-series: considerable
noise can be expected to result from both source flux variations and the \swift\ uneven
sampling of the light curve, and a search restricted around the \xmm\ period would have
involved the risk of selecting a spurious signal, in the case the true signal was embedded in
a high level of non-white noise. As expected, significant noise is present, but a very
prominent peak at 24\,031(2) s (the quoted uncertainty indicates the Fourier period
resolution) stands out well above the noise level, with a Leahy-normalised power \citep{leahy83} of 425
(Fig.~\ref{powspec}). While the non-white noise does not affect the frequency of a real
signal, it alters the statistical properties of a time-series; following the recipes of
\citet{israel96} we estimate that, after taking into account the number of frequencies
searched, the probability of having a signal this strong by chance coincidence is lower than
$3\times10^{-9}$ (that is a detection at a higher than 5.9$\sigma$ confidence level).
Moreover, the signal is consistent with the periodicity measured by \xmm. We also note that
no other periodicity with significance higher than 3$\sigma$ was found up to $\sim$1 year.

We folded the \swift\ data, as well as those of the observations in Table~\ref{obs-log}, on
the period $P=24\,031$ s. We obtained very significant pulse profiles; those of the data taken
during the quiescent state of \src\ (\swift, \xmm, \cxo/HRC-S and ACIS-I) are single-peaked (and well-modelled
 by two or three sine functions with the periods fixed at the fundamental period and
higher harmonics, with phases and amplitudes free to vary), while that from the \cxo/ACIS-S
observation shows two asymmetric peaks per cycle, both exhibiting two sub-peaks. In order to
obtain a refined ephemeris, we studied the phase evolution through the epoch-folded data by
means of an iterative phase-fitting technique (see e.g. \citealt{dallosso03}). Given the
variability of the pulse shape, we did not make use of a pulse template to cross-correlate
with, but we inferred the phase of the modulation by fitting each individual folded profile
with the fundamental plus three higher harmonics.\footnote{This is the minimum number of
harmonics necessary to properly fit all the folded profiles, including the multi-peaked
\cxo/ACIS-S one. The timing analysis described in the following was performed also with a
slightly different approach: the phases of the individual folded profiles were derived by a
fit with a variable number of harmonics, determined on a case-by-case basis by requesting that
the addition of a further (higher) harmonic is not statistically significant (by means of a
Fisher-test) with respect to the null hypothesis; the results are essentially identical.}

\begin{figure} \centering
\resizebox{\hsize}{!}{\includegraphics[angle=0]{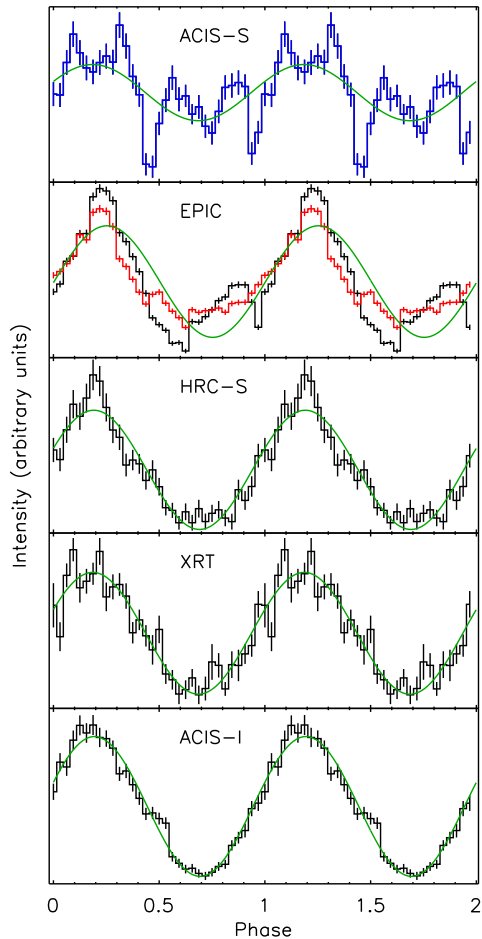}}
\caption{\label{pprofiles} 32-bin epoch-folded pulse profiles of \src\
obtained from different instruments (as indicated in each panel). The blue line distinguishes
the \cxo/ACIS-S data, which were used in deriving solution B  but not for solution A
(see Section~\ref{analysis} for details). In red we over-plotted the \xmm/EPIC data before
the filtering for proton flares. The fundamental harmonic of the pulse profile
is shown in green in each panel (in the EPIC panel it refers to the filtered data).}
\end{figure}

As a first step, we fit the phases of fundamental harmonic obtained from the \swift\ data
divided into four segments of approximately equal length and from the \xmm\ and \cxo\ HRC-S
and ACIS-I observations. At this stage, the \cxo/ACIS-S observation was left out, because of its very
different pulse profile. The time-evolution of the phase can be followed unambiguously
throughout all the data and described with a linear relation of the form
$\phi=\phi_0+2\pi(t-t_0)/P$. We assumed the start of the \swift\ monitoring, MJD 55\,804.0, as
the reference epoch $t_0$ and the fit ($\chi^2_\nu=0.88$ for 5 dof) gives $P_{\mathrm{A}}=24\,030.42(2)$ s
(1$\sigma$ uncertainty, valid over the range MJD 53\,605--55\,632). We designate this
rotational ephemeris `solution A'. A quadratic term
$-\pi(t-t_0)^2\dot{P}/P^2$, which would reflect the presence of a period derivative
($\dot{P}$), is not required. This implies an upper limit on the period derivative of \src\ of
$|\dot{P}_{\mathrm{A}}|<3.3\times10^{-9}$ s s$^{-1}$ (3$\sigma$ confidence level).
In Fig.~\ref{pprofiles} we show the
epoch-folded pulse profiles (including the \cxo/ACIS-S one) obtained using this solution. We
note that the \xmm\ observation was affected by rather intense proton flares.
The removal of the intervals of flaring background, because of the few cycles contained in
the observation,
significantly affects the pulse profiles. As a check of the robustness of our results, we
repeated the timing analysis using the unfiltered EPIC data and we obtained virtually
identical results (both filtered and unfiltered profiles are plotted in Fig.~\ref{pprofiles}).

Using solution A, we are able to predict for \src\ the phase of the
fundamental harmonic at the epoch of the \cxo/ACIS-S observation within
$\pm$0.03 cycles (at 3$\sigma$). The phase of the fundamental harmonic \emph{measured}
in the Chandra/ACIS-S data nicely dovetails with the predicted value.
Thus we derived a new coherent timing solution, which we denote with `B', including the
\cxo/ACIS-S data and therefore valid over the range MJD 52\,336--55\,632.
Again, the phases of the fundamental harmonic can be fit with a
linear relation ($\chi^2_\nu=0.77$ for 6 dof) which yields $P_{\mathrm{B}}=24\,030.42(2)$ s
(1$\sigma$ uncertainty; epoch MJD 55\,804.0).
While the best-fitting period is equal to that of solution A, the limit on the period derivative
is slightly more constraining:
$|\dot{P}_{\mathrm{B}}|<1.6\times10^{-9}$ s s$^{-1}$ (3$\sigma$ confidence level).

\section{Discussion} We presented the analysis of the first five years
(2006 April--2011 April) of the \swift/XRT
monitoring of the enigmatic X-ray source \src\ at the centre of the SNR \snr.
During this time span, the source remained in a quiescent state, with an average observed
1--10 keV flux of $\sim$$1.7\times10^{-12}$ \flux, 20--30 times lower than the historical
maximum attained in the 1999--2000 outburst ($\sim$$5\times10^{-11}$ \flux;
\citealt{garmire00}). The timing study of the \swift\ data yielded an accurate measure
of the modulation period of \src\ which allowed us to phase-connect the XRT data with archival
\cxo\ and \xmm\ observations. We derived two timing solutions, labelled A and B, both consistent
with the same constant period $P_{\mathrm{A}}=P_{\mathrm{B}}=24\,030.42(2)$ s but with somewhat different
upper limits on the period derivative. Solution A ($|\dot{P}_{\mathrm{A}}|<3.3\times10^{-9}$ s s$^{-1}$;
MJD 53\,605--55\,632) is based on the single-peaked pulse profiles observed while \src\ was in
quiescence. Solution B ($|\dot{P}_{\mathrm{B}}|<1.6\times10^{-9}$ s s$^{-1}$; MJD 52\,336--55\,632)
is a natural extension of solution A including the multi-peaked \cxo/ACIS-S profile obtained on
2002 March 03, when \src\ was in a rather bright state \citep{sanwal02}.

The phase-coherent timing technique employed in our timing study of \src\ closely parallels
the well-tested (virtually all systematics are under control) procedures used for X-ray bright
spin-powered pulsars.
% For such objects it is implicitly assumed that the harmonics reflect a common phase
% delay that traces the rotational phase $\phi$ of the star\footnote{In some cases the
% assumption is reasonable also for the orbital phase of a binary system.}
% ($\Delta\phi_k\approx\Delta\phi$ for any $k$, where $k$ indicates the $k$-th harmonic, i.e.
% $k$ times the fundamental frequency) and that any variabilty is due to Poisson noise.
%
For such objects it is implicitly assumed that the pulse profile does not intrinsically change in
time, so that, when a Fourier decomposition of the pulse is introduced, neither the fundamental
nor the higher harmonics evolve. In such a picture all the variability is due to Poisson noise and the phase
evolution can be tracked by matching the pulse profile at any given epoch always with the same template.

In the case of \src\ there are clear indications that the pulse shape changes in time so that
the previous assumptions are not valid. This forced us to abandon the standard
template-matching analysis and follow instead the phase of the
fundamental (first) harmonic.\footnote{The large phase
uncertainties do not allow a similar study of the higher harmonics (which are not even always detectable).}
%The limit of this procedure lies in the possibility that the first
%harmonic might have less fundamental physical meaning than the higher ones have.\footnote{For
%example, even if the stability of the phase evolution of the fundamental harmonic over several
%years suggests that this is not the case for \src, the fundamental harmonic could be linked to
%a more intense but less stable emission component.}
The main justification for such an approach is that usually
(moderate) pulse shape changes are associated with higher
harmonics while the fundamental is stable. Even if this
seems to be the case for \src, given the stability of the phase
 of the fundamental harmonic over several years, we
stress that it does not have to hold in general (see, e.g., \citealt{hartman08}).

In the following we discuss the implications of our newly derived upper limit 
$\vert\dot P\vert < 1.6\times 10^{-9}$ s s$^{-1}$ on the models that have been 
proposed so far for \src. We remark that all the ensuing considerations are based 
on the assumption (discussed above) that the fundamental harmonic is a good tracer 
of the timing behaviour of the source.

Although the nature of \src\ is still an open issue, most
interpretations favour the neutron star scenario, in which the
star is either isolated \cite[][]{deluca06,li07} or in a binary
system with a low-mass companion
\cite[][]{pizzolato08,bhadkamkar09}. While the model by
\cite{bhadkamkar09} is based on a fast-spinning, moderately
magnetised neutron star in a 6.67 hr eccentric orbit, in all the
other cases an ultra-magnetised neutrons star ($B\approx 10^{15}\
{\rm G}$) is required to explain the very long period of \src\ and
the observed X-ray periodicity is related to the star spin period
(in the binary scenario of \citealt{pizzolato08} the latter may or
may not coincide with the orbital period). Actually, as already
noted by \cite{deluca06}, magneto-dipolar braking alone is not
enough to explain the present value of $P$ even invoking a
magnetar, and spin-down by the interaction of the star
magnetosphere with a (residual) disc is also necessary for an
isolated object. Spin-down to $P\simeq 6.67\ {\rm hr}$ in a time
$\sim$2 kyr can be achieved if the initial period is peculiarly
long ($\ga$300 ms, \citealt{deluca06}; magnetars are in fact believed to be born
with ms periods, \citealt{thompson93}). Using different assumptions
on the termination radius of the disc \cite[see][]{eksialpar05},
\cite{li07} showed that the same result can be recovered also for
more conventional values of the initial period ($\approx$10 ms).

The current spin-down rate expected in the binary magnetar model
is very small since the equilibrium period is reached well in
advance of $\sim$1000  yr, unless the synchronisation time is very
short ($\sim$10 yr) and there is no mass transfer in the system
\cite[][]{pizzolato08}. If \src\ is an object of this kind our
upper limit on $\dot P$ is completely non-constraining for the
model.

On the other hand, if \src\ is an isolated magnetar
surrounded by a fossil disc it must be a quite rare system.
According to the Monte Carlo simulations of \cite{li07}, the
fraction of objects of this type with periods $\ga$100 s at an age
of $2.5$ kyr is only $\sim$1\%. The large majority of stars are
much faster rotators ($P\sim\ 1$--10 s) in the ejector stage
(typically identified with SGRs/AXPs) while ultra-slow systems are
in the propeller/accretor phase.\footnote{Although previous
figures are model-dependent the conclusion that
propellers/accretors are a tiny minority appears to be robust.}

\begin{figure} \resizebox{\hsize}{!}{\includegraphics[angle=0]{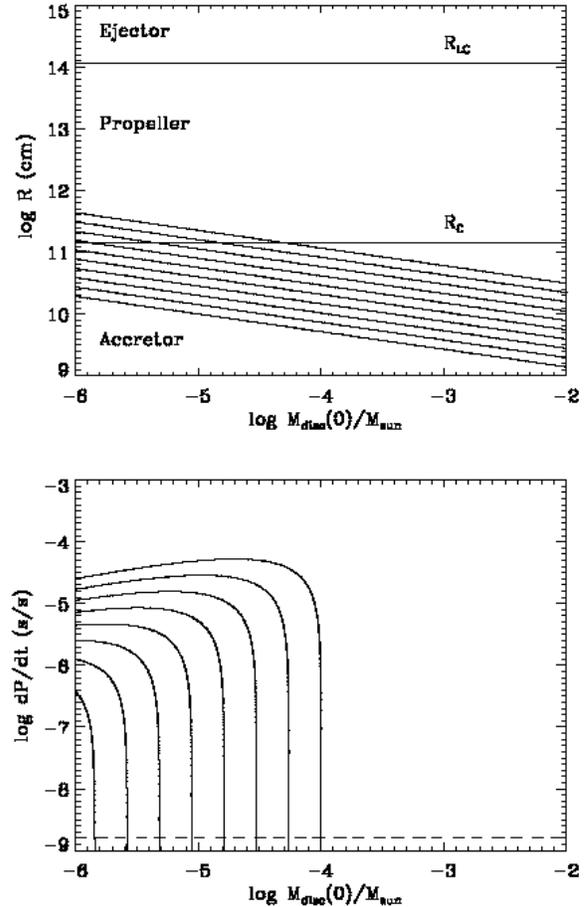}}
\caption{\label{pdot} Upper panel: the Alfv\'en radius as a
function of the initial disc mass $M_D(0)$ for different values of
the magnetic field in the range $13.5\leq \log B\, (\rm G)\le 16$
(the field increases as the curve moves upwards); the light
cylinder radius $R_{LC}$ and the corotation radius $R_C$ are also
shown. Lower panel: the period derivative as a function of
$M_D(0)$ for those values of $B$ for which the source is currently
in the propeller stage; the dashed line marks our derived upper limit
on $\dot P$. Here the period and age have been fixed to those inferred
for \src\ (see text).}
\end{figure}

The spin-down rate of a neutron star which interacts with a fossil
disc in the propeller stage is given by $\dot P \sim -\dot
MR_{in}^2[\Omega_K(R_{in})-2\pi/P]P^2/(\pi I)$ where
$R_{in}=0.5[B^4R^{12}/(8GM\dot M^2)]^{1/7}$ is the Alfv\'en
radius, $M\sim 1.4M_\odot$ and $R\sim 10^6\ {\rm cm}$ are the star
mass and radius ($I\sim 10^{45}\ {\rm g\,cm}^2$ is the moment of
inertia), $B$ is the surface magnetic field, $\Omega_K$ is the
Keplerian angular velocity and $\dot M\propto t^{-1.25}$ is the
mass loss rate from the disc \cite[][and references
therein]{li07}.\footnote{We remark that no complete theory of the
interaction of a disc with a magnetised neutron star exists.
Following \cite{li07} we adopted the `efficient' form of the
propeller torque \cite[see e.g.][for a comparison of different
expressions of the torques]{franci02}.} If \src\ is
currently in the propeller phase, $\dot P$ can be derived
from the previous expressions with $P\sim 24$ ks and $t\sim 2.5$
kyr, as a function of the disc initial mass $M_D(0)$ and of $B$.
Results are shown in Fig.~\ref{pdot} (lower panel) for
$10^{-6} < M_D(0)/M_\odot < 10^{-2}$ and $3.2\times 10^{13}\ {\rm
G} < B < 10^{16}\ {\rm G}$; the Alfv\'en radius as a function of
$M_D(0)$ and $B$ is plotted in the
upper panel. Comparison of $R_{in}$ with the light cylinder
radius, $R_{LC}=cP/(2\pi)$, and the co-rotation radius,
$R_{C}=[GMP^2/(4\pi^2)]^{1/3}$ shows that the star is currently 
in the propeller phase, characterised by $R_C<R_{in}<R_{LC}$, only if $B\ga 10^{15}\ \rm G$. 
Lower values of the field or large enough (depending on $B$) initial disc mass would
result in \src\ being in the accretor stage. The resulting values of $\dot P$ 
are orders of magnitude above the upper limit we reported (as a reference, the
case discussed in \citealt{deluca06} has $\dot P\sim 10^{-7}$ s
s$^{-1}$), with the only exception of an extremely narrow range of
$M_D(0)$ for each $B$. Although not all the values of $M_D(0)$
and $B$ we considered will produce the correct period at an age of
2.5 kyr (this depends also on both the initial period and the
angle between the magnetic and spin axes, see again
\citealt{li07}), our conclusion is that if \src\ is an isolated
magnetar, and hence it must be either a propeller or an accretor
at present, the typical large expected values of $\dot P\gg
3\times 10^{-9}$ s s $^{-1}$ makes the propeller option rather
unlikely. Assessing the reliability of the accretor scenario
requires a more thorough analysis. Here we just note that if the
same expression of the torque remains valid during the accretion
phase, the spin-up rate is also largely in excess of our upper
limit on $\dot P$.

\section*{Acknowledgments}
We thank the referee, David Helfand, for his constructive and valuable comments.
This research is based on data and software provided by the
NASA/GSFC's High Energy Astrophysics Science Archive Research Center (HEASARC), the \cxo\
X-ray Center (CXC, operated for NASA by SAO), and the ESA's \xmm\ Science Archive (XSA). PE
acknowledges financial support from the Autonomous Region of Sardinia through a research grant
under the program PO Sardegna FSE 2007--2013, L.R. 7/2007 ``Promoting scientific research and
innovation technology in Sardinia''.  This work was partially supported by the ASI/INAF
contract I/009/10/0.

%\bibliographystyle{mn2e}
%\bibliography{biblio}

\bsp

\label{lastpage}

\end{document}